\documentclass[multphys,vecphys]{svmult}

\usepackage{makeidx}         
\usepackage{graphicx}        
\usepackage{multicol}        
\usepackage[bottom]{footmisc}

\makeindex             

\begin{document}

\title*{Population of dynamically formed triples in dense stellar systems}
\author{Natalia Ivanova}
\institute{Physics and Astronomy, Northwestern University, 2145 Sheridan Rd,
Evanston, IL 60208 USA 
\texttt{nata@northwestern.edu}}
%
%
\maketitle
\begin{abstract}
In dense stellar systems, frequent dynamical interactions between
binaries lead to the formation of multiple systems.
In this contribution we discuss the dynamical formation
of hierarchically stable triples: the formation rate,
main characteristics of dynamically formed population of triples
and the impact of the triples formation on the population of close binaries.
In particular, we estimate how much
the population of blue stragglers and compact binaries could be affected.
\end{abstract}

\section{Introduction}
In globular clusters, the most plausible
way for the dynamical formation of 
hierarchical triples is via binary-binary encounters.
As numerical scattering calculations show, the probability
of triple formation is quite substantial 
and for equal masses, a hierarchical triple
is formed in roughly 50\% of all encounters (Heggie \& Hut 2003).
The probability is reduced by only a factor of a few when original semi-major
axis of binaries are about equal (Mikkola 1984).
The consideration of stars as non-point masses
can decrease this probability further, as the physical collisions
enhance strongly the destruction of close binaries during binary-binary encounters (Fregeau et al. 2004).
The triple formation have often been noticed in numerical simulations of dense stellar systems
using $N-$body codes (e.g., McMillan et al. 1991, Hurley et al. 2005),
however so far there has been no attempt to study in detail the population of triple 
systems as well as their effect on the close binaries and blue stragglers.
In this contribution we report the preliminary results of
our triples population study.

\section{Method and assumptions}

We use a {\it Monte Carlo\/} method described in detail in Ivanova et al. 2005.
This method assumes a static cluster background and that
all relevant dynamical parameters are kept constant throughout dynamical simulation.
In particular, the cluster model we consider here 
has central density $n_{\rm c}=10^5$ [pc$^{-3}$], velocity dispersion $\sigma=10$ [km/s], 
escape velocity $v_{\rm esc}$ [km/s] and half-mass relaxation time $t_{\rm rh}=10^9$ [yr].
The code takes into account such 
important dynamical processes as mass segregation and evaporation, 
recoil, physical collisions, tidal captures, and binary--single
and binary--binary encounters.
For dynamical encounters that involve binaries we use
{\tt Fewbody}, a numerical toolkit for direct $N$-body integrations
of small-$N$ gravitational dynamics (Fregeau et al. 2004).
This toolkit is particularly suited to automatically recognize 
a hierarchical triple  (formed via an encounter) using the stability criterion 
from Mardling \& Aarseth (2001).
In order to get large statistics on triples formation rate, 
we start with $1.25 \times 10^6$ stars, 100\% are in primordial binaries;
the modeled cluster has mass $\sim 250,000 M_\odot$ at 10 Gyr and
the core mass is 10-20\% of the total cluster mass.
This cluster model represents well a ``typical'' globular cluster.

There is no developed population synthesis methods for triples evolution 
and as a result we can not keep the triples once they were formed. 
In our standard runs, we break a triple conserving the energy:
the energy required to eject the outer companion is acquired from shrinking of the inner binary orbit.  
The outer companion is released unless the inner binary merges during shrinkage.
In the latter case the inner system is allowed to merge and the outer companion is kept 
at its new, wider orbit to form the final binary system.
This treatment prevents the possible eccentricity increase via the Kozai mechanism (Kozai 1962),
which causes large variations in the eccentricity and inclination of the star orbits
and could drive the inner binary of the triple system to merge before next
interaction with other stars.
To check this effect on the binary population, we compare two cluster models, with the completely 
the same initial population of $5\times 10^5$ stars.
In one model we use our standard treatment for triples breaking.
In the second model we compare the Kozai time-scale $\tau_{\rm Koz}$ (taken as in Innanen et al. 1997)
and the collision time-scale $\tau_{\rm coll}$ and
inner binaries in the formed triples are merged if $\tau_{\rm Koz} < \tau_{\rm coll}$ (we define
these triples as Kozai triples).
We expect that some of the triples can also have a secular eccentricity evolution (see e.g. Ford et al. 2000),
but we neglect this possibility.
                                             
\section{Numerical results on dynamically formed triples}

\subsection{Formation rates}

From our standard model, we find that  a ``typical'' cluster has about 5000 hierarchically
stable triples formed in its core throughout its evolution.
As our triples are formed via-binary-binary encounters, the resulting
formation rate intrinsically depends on the binary fraction, the binary cross-section 
and the total number of binaries.
In the result, the obtained formation rate can be written as:

\begin{equation}
\Delta N_{\rm tr} / N_{\rm bin} = 0.05 f_{\rm bin}  \langle m_{\rm b} \rangle \langle a \rangle {\rm \ per\ Gyr}.
\end{equation}
 Here $f_{\rm bin}$ is the binary fraction, $ \langle m_{\rm b} \rangle$
is the average binary mass and $\langle a \rangle$ is the average
binary separation. 
In particular, at 10 Gyr  $ \langle m_{\rm b} \rangle \approx 1 M_\odot$,  $\langle a \rangle \approx 10 R_\odot$
and  $f_{\rm bin} \approx 10\%$. 
The formation rate at 10 Gyr is therefore such that 
as many as 5\% of all core binaries have sucesfully 
participated in the formation of hierachically stable triples during 1 Gyr.
We stress that the expression above is fitted to numerical simulations; it does not include directly the expected dependence 
on the core number density and velocity dispersion 
because only one set of these parameters was used in simulations.

The formation rate of triples can also be written as the function of the cluster age (for ages $> 1$~Gyr):

\begin{equation}
\Delta N_{\rm tr} = 600\ T_{\rm 9}^{-1/3} {\rm \ per\ Gyr}.
\end{equation}
Here $T_{\rm 9}$ is the cluster age in Gyr.

\subsection{Masses, orbital periods and eccentricities}

\begin{figure}
\centering
\includegraphics[height=5.8cm]{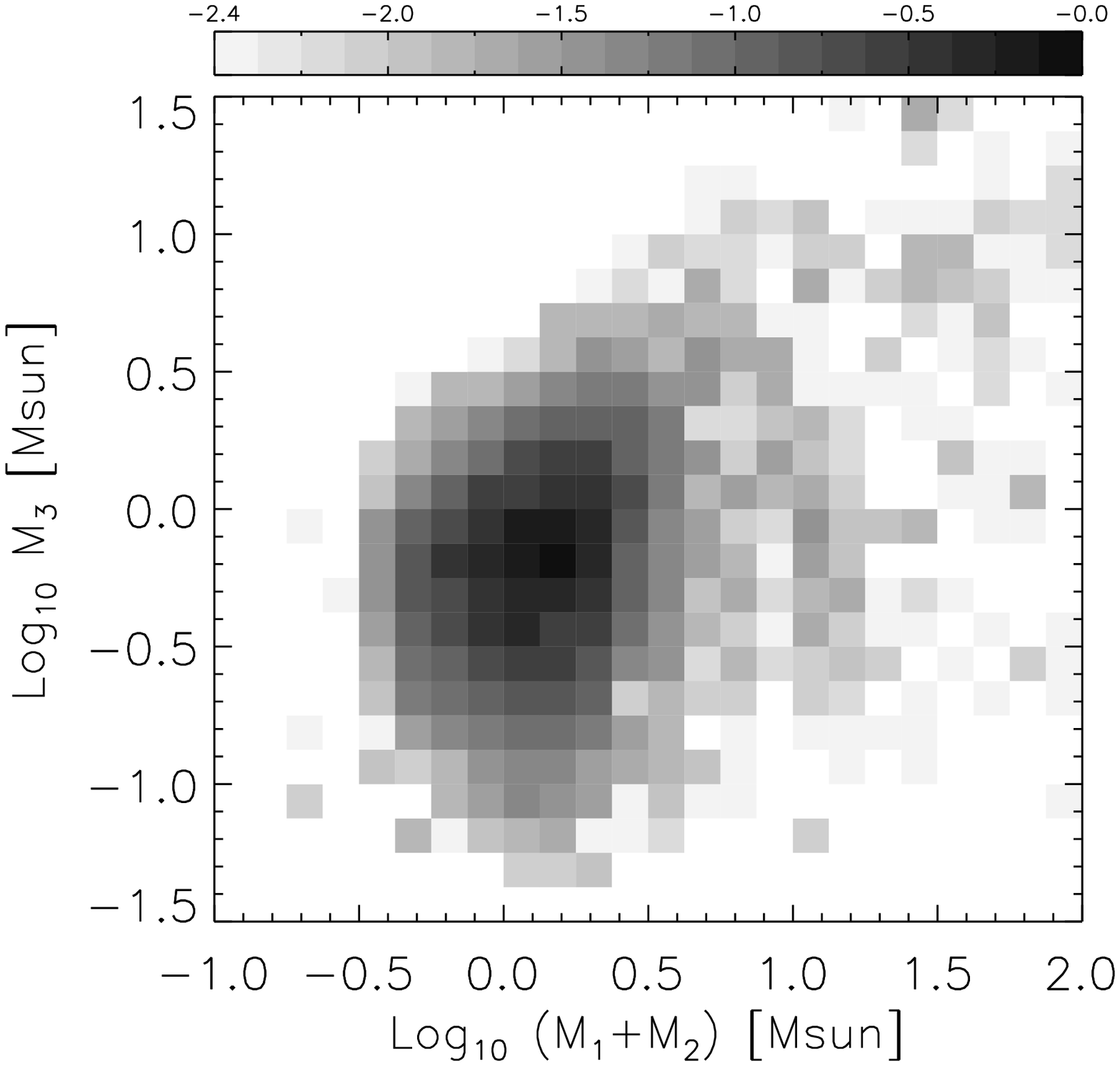}\includegraphics[height=5.8cm]{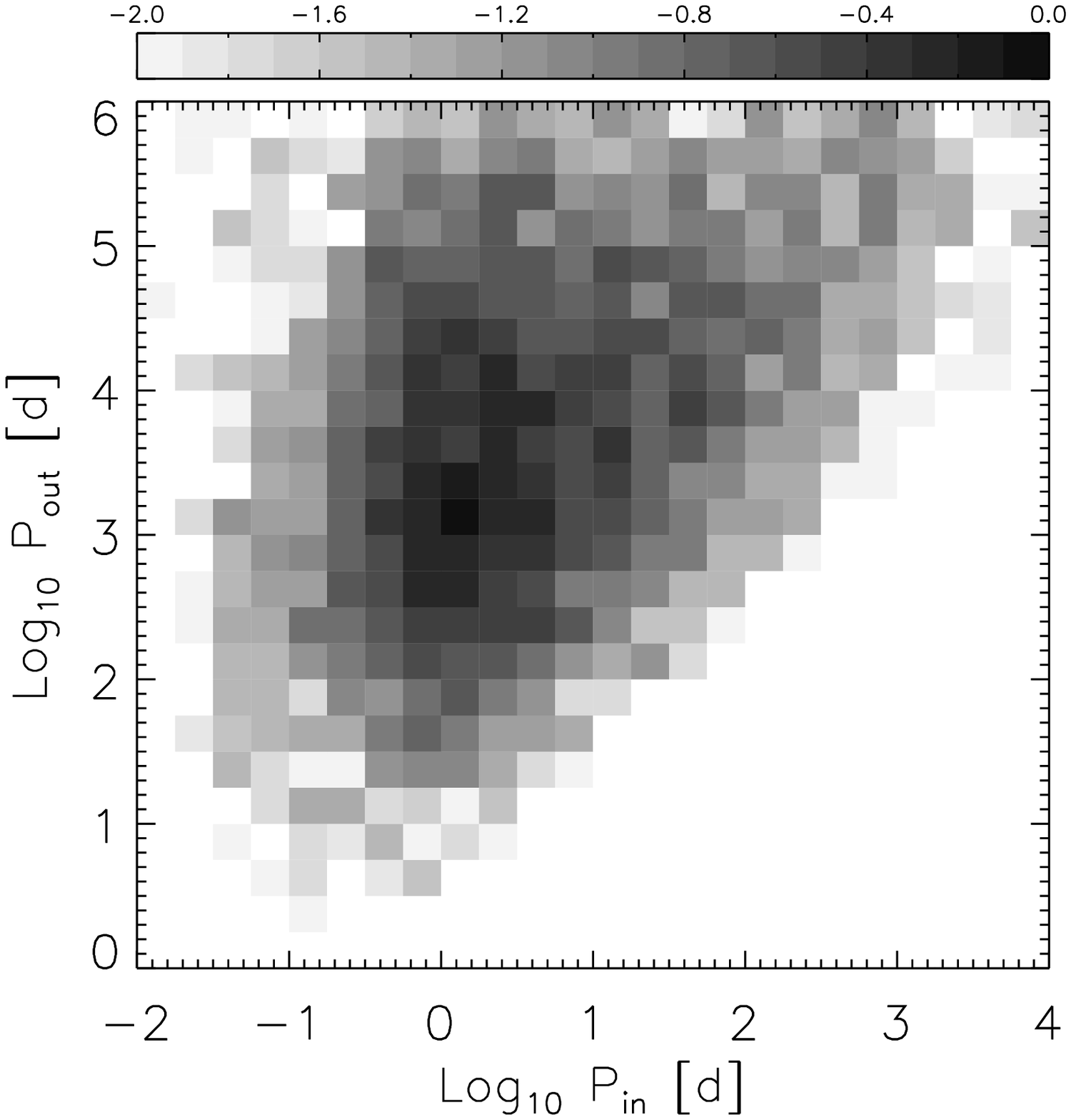}
\includegraphics[height=5.8cm]{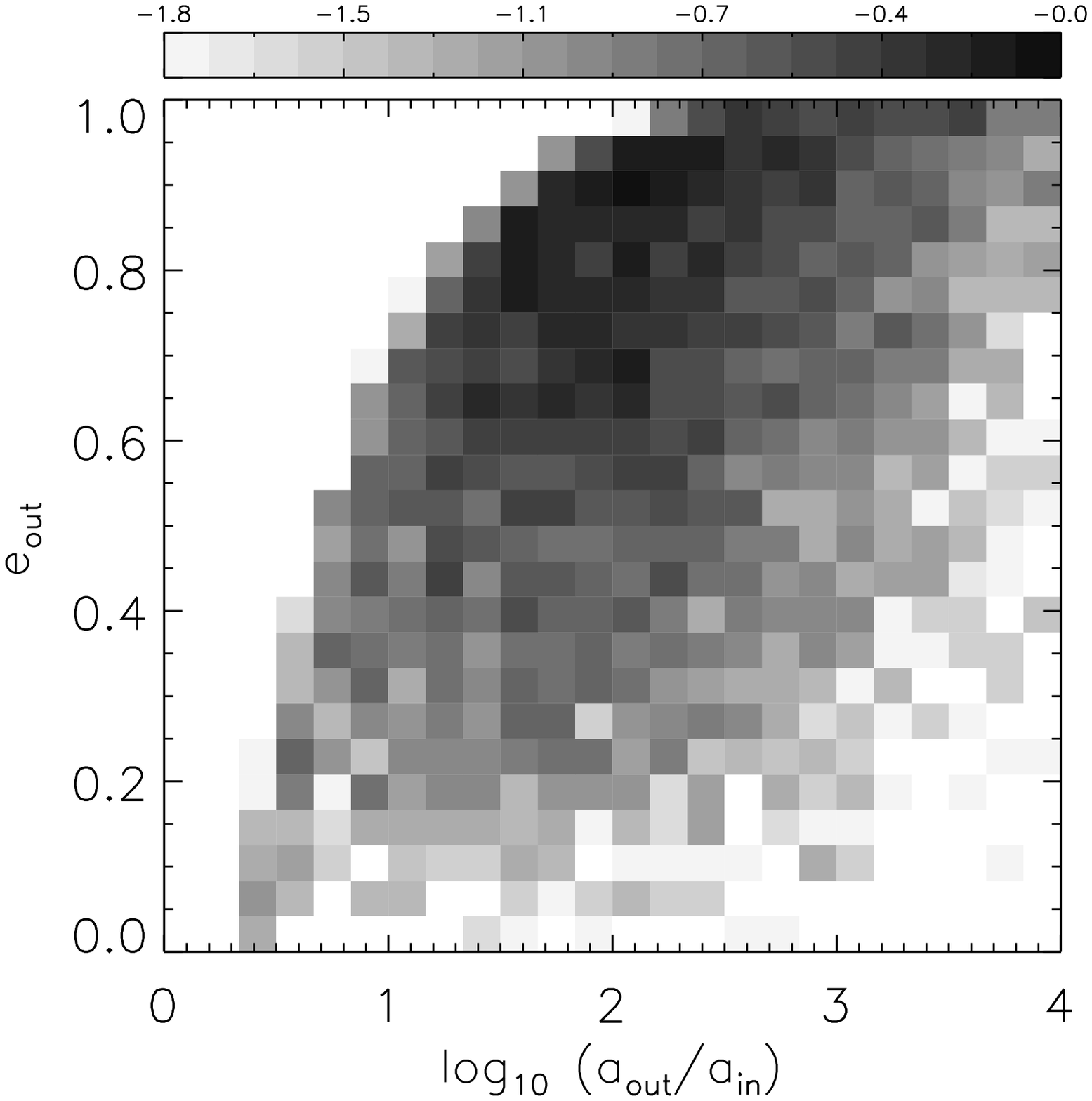}
\caption{The distributions of masses of inner binaries and the outer companions (the left panel),
the inner and the outer orbital periods  (the right panel) and the outer companion eccentricities
versus the orbital separation ratio (the bottom panel) in the dynamically formed hierarchically stable triples.
The colors correspond to the logarithm of the probability density.}
\end{figure}

In Fig.~(1) we show the distributions of the triples companions masses (the mass of the inner binary and 
of the outer companion), the inner and the outer orbital periods and the outer eccentricities, for all 
dynamically formed triples throughout the entire cluster evolution.
The inner eccentricities did not show strong correlation with any other parameters and are
distributed rather flatly.
A ``typical'' triple has the mass ratio $M_3/(M_1+M_2) \approx 0.5 \pm0.1$, the total mass if
the inner binary is $M_1+M_2 \approx 1.3\pm 0.3 M_\odot$ (such a binary, if merges, will likely produce a blue straggler),
$P_{\rm in}\approx 1$ day, high period ratio $P_{\rm out}/P_{\rm in}\approx 1000 $, the ratio of the orbital separations
$a_{\rm out}/a_{\rm in}\approx 100$ and very high outer eccentricity, $e_{\rm out}\approx 0.95\pm0.05$.

\subsection{Hardness and the Kozai mechanism}

Only these  triples that have the
binding energy of the inner binary with the outer companion greater than a kinetic energy of an average
object in the field are stable against the following
dynamical encounters (the ratio of these energies is the triple hardness, $\eta$). 
We find that 45\% of all triples have $\eta > 1$ and only 7\%
of all triples have $\eta> 10$. 
In our numerical simulation we find that for binaries, to likely survive
subsequent dynamical encounters, a hardness should be about few times larger than 1. 
Therefore we can assume that most of the formed triples
can be easily destroyed in their subsequent evolution in the dense core.
However, we find also that about a third of all triples are the Kozai triples. 
The probability that a triple is affected by 
Kozai mechanism does not correlate with the triple hardness, the orbital periods or the eccentricities. 
In the result, a significant fraction of all triples  
can evolve not as our triples breaking treatment predicts.

\subsection{Population of Kozai triples}

\begin{figure}
\centering
\includegraphics[height=5.8cm]{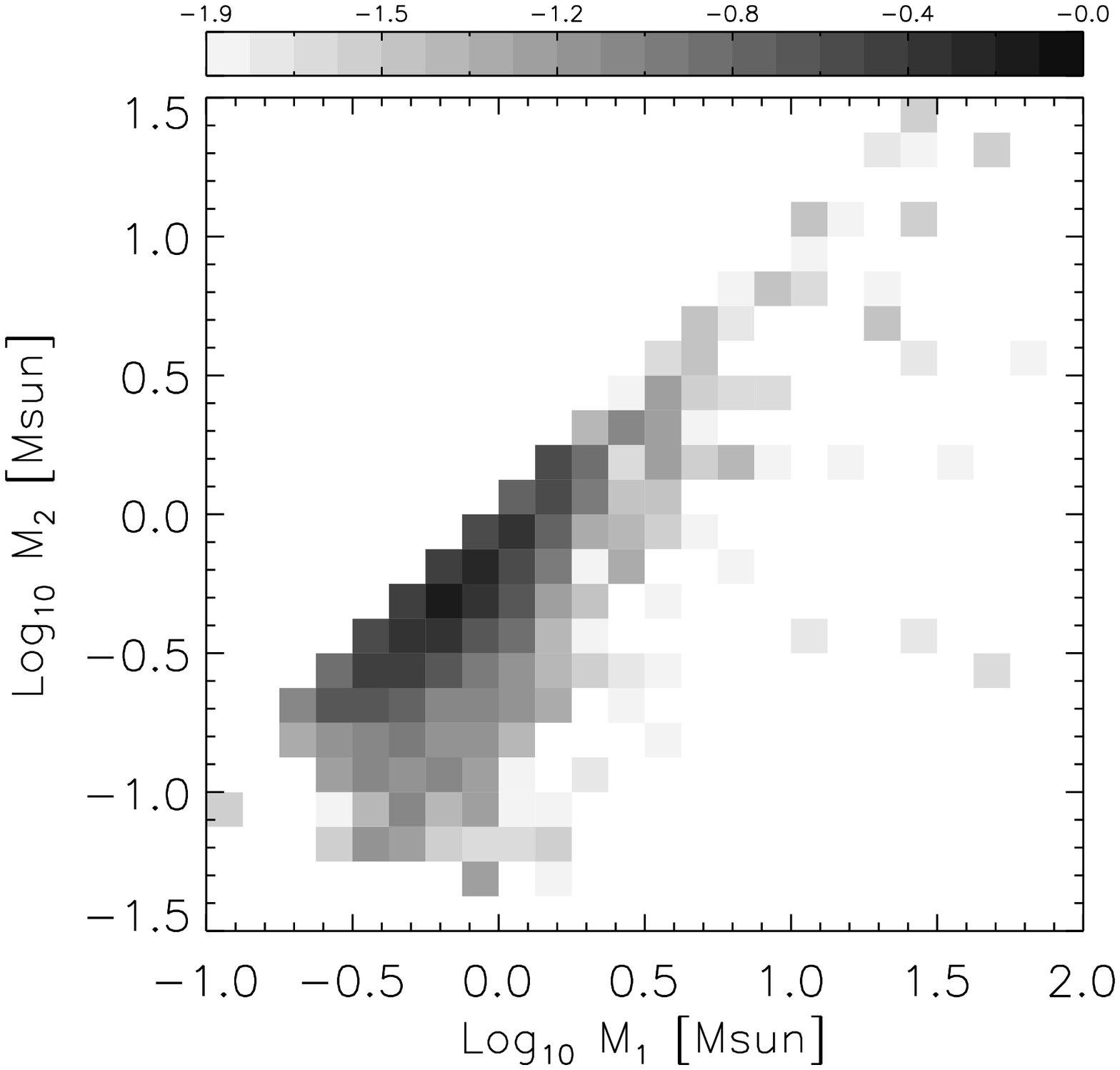}\includegraphics[height=5.8cm]{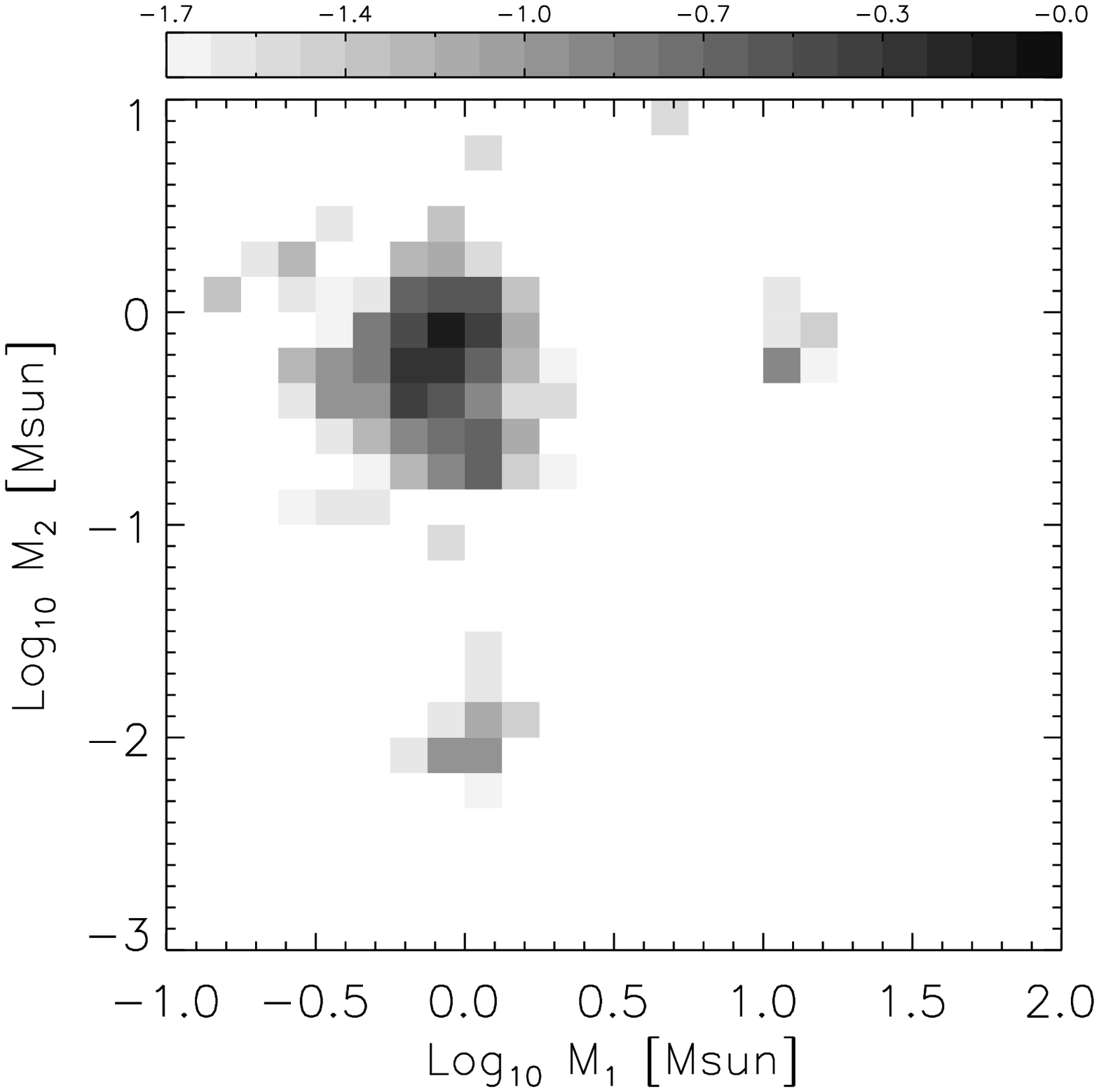}
\caption{The population of inner binaries in Kozai triples: main sequence-main sequence binaries (the left panel) and
and inner binaries with a compact object (the right panel). 
The colors correspond to the logarithm of the probability density, $M_1$ 
is the mass of more massive companion in the case of main sequence-main sequence binary or the mass of a compact companion.}
\end{figure}

About 60\% of all triples have main sequence-main sequence inner binary and 30\% of them are
Kozai triples. 
A typical Kozai main sequence-main sequence binary has the total mass of the inner 
binary of $1.3\pm0.2$ $M_\odot$ (see also Fig.~2).
If Kozai mechanism leads to a merger,
we find that these triples can provide $10\%$ of all ever created blue stragglers.

About 40\% of all triples have inner binary with a compact object and 35\% 
(40\% for white dwarf-white dwarf inner binaries) of them are Kozai triples.
A typical white dwarf-main sequence binary affected by Kozai is 0.8 $M_\odot$ 
white dwarf and 0.8 $M_\odot$ main sequence star.
It is not clear if the Kozai mechanism will lead to the merger or to the
start of the mass transfer and create therefore a cataclysms variable.
The formation rate per Gyr (at ~ 10 Gyr age) corresponds to about 25-50\%
of present at this age cataclysms variables.

\subsection{Comparison of the close binaries population and the triples population}

\begin{table}
\centering
\caption{The stellar population of close binaries and triples in the cluster core at the age of 10 Gyr. MS is for a main sequence star,
RG is for a red giant, WD is for a white dwarf and NS is for a neutron star.}
\begin{tabular}{lllllll}
\noalign{\smallskip}\hline
\noalign{\smallskip}\       &  Binaries & &   Triples,  & & Triples, \\
\noalign{\smallskip}\       & \ & \ & inner & binaries \ \ &  outer companions\\
\noalign{\smallskip}\hline
\noalign{\smallskip} \ & MS\phantom{xxxxxx} & WD\phantom{xxxxxx}  & MS\phantom{xxxxxx} & WD\phantom{xxxxxx} &\phantom{xxxxxx} \\
\noalign{\smallskip}
\noalign{\smallskip}\hline\noalign{\smallskip}
MS & 79\% & 13\%  & 60\% & 20\%  & 79\%  \\
RG & 0.7\% & 0.3\% & 2\% & 1\% & 0.7\%\\
WD &  & 7\%  & & 15\% & 20\%\\
NS & 0.3\% & 0.3\%  & 0.6\%& 0.7\% & 0.3\% \\
\noalign{\smallskip}\hline
\end{tabular}
\end{table}

In Table~1 we provide the complete statistics for the stellar population of close binaries
and triples in the cluster core at the age of 10 Gyr. We find that the inner binary of a triple is more likely to
contain a compact object than a core binary at 10 Gyr. Outer star follows the binaries populations.

\subsection{Comparison of models with and without Kozai triples mergers.}

We did not find that different treatments of Kozai triples (our standard breaking of triples
based on the energy balance or the enforced merge of the inner binary if the triples is a Kozai triple) 
lead to  significantly different results for the binary fractions (the difference is less than 1\%).
Some difference is noticed for the distribution of binary periods
of core binaries with a white dwarf companion: in the model where Kozai triples, once formed,
had their inner binaries  merged, the relative number of binaries with the periods less than 0.3 day
is smaller than in the model where Kozai triples were treated as usual, and for binaries
with periods between 0.3 and 3 days the situation is reversed.

\section*{Acknowledgments}

The author thanks John Fregeau for the useful discussion on triples stability
and acknowledges the support by a {\em Chandra}  Theory Award.

%
%
%
%
%

%

\begin{thebibliography}{99.}
%
%
%


 
\bibitem{2000ApJ...535..385F}  E.~B. Ford, B. Kozinsky, 
 F.~.A. Rasio: Astroph. Journal \textbf{535}, 385 (2000)  
\bibitem{2004MNRAS.352....1F}  J.~M. Fregeau, P. Cheung, 
S.~F. Portegies Zwart et al.: MNRAS \textbf{352}, 1 (2004)
\bibitem{2003gmbp.book.....H} D. Heggie, P. Hut:
\textit{The Gravitational Million-Body Problem: A Multidisciplinary Approach 
to Star Cluster Dynamics} (Cambridge  University Press 2003)  
\bibitem{2005astro.ph..7239H} J.~R. Hurley, O.~R. Pols, 
S.~J. Aarseth, S.~J. at al: ArXiv Astrophysics e-prints, 
arXiv:astro-ph/0507239 (2005)
\bibitem{1997AJ....113.1915I} K.~.A. Innanen, J.~Q. Zheng, 
S. Mikkola et al: Astronomical Journal \textbf{113}, 1915 (1997)
\bibitem{2005MNRAS.358..572I} N. Ivanova, K. Belczynski, J.~M. Fregeau et al: MNRAS  \textbf{358}, 572 (2005)
\bibitem{k62} Y. Kozai:Astronomical Journal  \textbf{67}, 591 (1962)
\bibitem{maraar01} R.~A Mardling, S.~J. Aarseth:MNRAS 2001 \textbf{321}, 398 (2001)
\bibitem{1984MNRAS.207..115M} S. Mikkola: MNRAS \textbf{207}, 115 (1984)
\bibitem{1991ApJ...372..111M} S. McMillan, P. Hut, J. Makino: Astroph. Journal \textbf{372}, 111 (1991)

\end{thebibliography}
%



\printindex
\end{document}